# TEXT MINING THROUGH LABEL INDUCTION GROUPING ALGORITHM BASED METHOD

**Gulshan Saleem[1], Nisar Ahmed[2], Usman Qamar[1]**

[1] Department of Computer Engineering, College of EME, National University of Sciences and Technology (NUST), H-12, Islamabad, Pakistan

[2] Department of Computer Engineering, University of Engineering & Technology, Lahore, Pakistan

Corresponding Author: gulshan.saleem14@ce.ceme.edu.pk

**ABSTRACT:** The main focus of information retrieval methods is to provide accurate and efficient results which are cost effective too. LINGO (Label Induction Grouping Algorithm) is a clustering algorithm which aims to provide search results in form of quality clusters but also have few limitations. In this paper, our focus is based on achieving results which are more meaningful and to improve the overall performance of the algorithm. LINGO works on two main steps; Cluster Label Induction by using Latent Semantic Indexing technique (LSI) and Cluster content discovery by using the Vector Space Model (VSM). As LINGO uses VSM in cluster content discovery, our task is to replace VSM with LSI for cluster content discovery and to analyze the feasibility of using LSI with Okapi BM25. The next task is to compare the results of a modified method with the LINGO original method. The research is applied on five different text based data sets to get more reliable results for every method. Research results show that LINGO produces 40-50% better results when using LSI for content Discovery. From theoretical evidences using Okapi BM25 for scoring method in LSI (LSI+Okapi BM25) for cluster content discovery instead of VSM, also results in better clusters generation in terms of scalability and performance when compares to both VSM and LSI's Results.

## INTRODUCTION

Today, we have a large amount of data from every domain and this is not the end but it has exponential property. Due to large amount of data it is not the simple task to get required knowledge and requires research attentions to get out of such situation. As the needs may vary from problem to problem so it is important to do problem analysis first, then to select a method to get solution from this world of data. There are a number of methods which are being used for information retrieval. Matching is the simple term which we normally use to group similar things. But in data mining, it is not as simple matching is further classified in various types. We have two main types learning methods; supervised Learning (Classification based) and unsupervised learning (Clustering based). Data Mining has sub domains as well and web mining is one of them. As in Figure 1, Web mining is based on analysis of data and further categorized as web structure mining, web usage mining and web content mining. Our research is basically based on Web content mining. We use the clustering method as information retrieval. Clustering is not itself an algorithm, but a method which we use to formulate algorithms for data/info retrieval formulate data. For this research, we used Label Induction Grouping Algorithm which is based on grouping of similar contents in a cluster. LINGO has capability to capture thematic threads in a search result. It simply discover groups of related documents and describe the subject of these groups in a human understandable way and combines several pre-existing techniques to put particular emphasis on meaningful cluster descriptions, in addition to discovering similarities among documents. As our corpus is based on text type dataset so the techniques used in LINGO are also text analysis techniques which are latent semantics indexing (LSI) and vector space models (VSM).

To do an analysis of the content of documents and queries available on the web is text analysis, which is also called as text mining. The aim of the text analysis method is a deduction of meaningful information from text data. It is not like matching does, but in depth we need patterns, associations and rules that are used in text data. Text mining is based on well organization of inputs given and then deduction, rating and interpretation of organized data. Some important tasks related to text mining are a categorization of text, concept mining/entity extraction, sentiment analysis, text summarization and text clustering.

The main contribution of this paper is based on the implementation of different modifications of LINGO algorithm and then to compare its results with the original method. We used LINGO as documents clustering algorithm on five different datasets. To improve its quality we uses latent semantic indexing in the second phase of cluster content discovery which means that assignment of documents to label formed at the first phase of cluster label induction. Furthermore, we also implemented LINGO with LSI+Okapi BM25 (Okapi BM25 for replacement of scoring method in LSI) using it in thesecond phase of cluster content discovery. And next effort is based on a comparison of results achieved in every method.

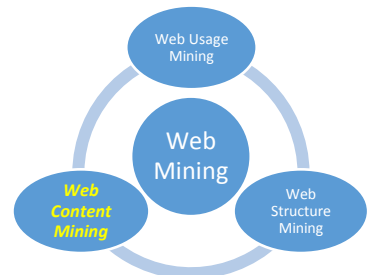

**Figure 1: web Mining**

**RELATED WORK**
To complete research, we did the thorough analysis of following techniques which are the main basis of these modifications in LINGO.

Lingo is a novel clustering algorithm designed by Osiński *et al.* [1]. Their work is based on detailed proposition and a description of LINGO algorithm. Authors have used the suffix arrays for phrase extraction and describe all the algebraic transformation of the term-document matrix. The authors provided the empirical evaluation method to support their techniques. Lingo is the inspiration form of the existing clustering techniques and a commercial system-Vivisimo. The authors also describe the future direction for the modification of the algorithm. It uses two well established and successful information retrieval techniques named as latent semantic indexing and the vector space model. Latent semantic indexing is utilized for the induction of clustering label, whereas the contents of the clusters are assigned by vector space model so for contents it uses lateral matching.

Chen et al [2] proposed a framework using latent semantic indexing for information retrieval. Authors have highlighted the issue with scalability of the system using large datasets. According to author information retrieval is not efficient as it cannot execute on large scale dataset in parallel fashion. To handle such data issues, they presented a framework which uses regularized latent semantic indexing with L2/2 regularization and non-negative constraints. By decomposing learning process into a chain of mutually independent sub-optimization problems which can be functioned in parallel so handles large-scale data. The proposed method is conductive and practical to information retrieval and relatedness computing. This framework uses a topic model which limits scalability of large data sets. And by combing it with L1/2 regularization and non-negative constraints, it became scalable and more effective too.

Mirzal [3] have worked with latent semantic indexing to attain good performance of an information retrieval system. LSI is an indexing method which index the terms appeared in related documents and weakens the influence of terms appears in unrelated documents. Latent semantic indexing is mainly based on singular vale decomposition method and the problem identified by the author is that it relies on choosing an appropriate decomposition rank. Author proposed a solution to this problem by using the fact of truncated SVD makes the document more connected. Author provided a Matrix completion algorithm which results in unique solution for each input. The proposed method is non-parametric and guarantees convergence. Within a cluster, LSI has the capability of strengthening connections, so by using this fact proposed algorithm uses cosine similarity and adjust weight based on the measures. Finally, convergence analysis of matrix completion algorithm resulted as the algorithm is convergent and gives a unique solution for each input matrix with cubic computational complexity. The Matrix completion algorithm provides a more practical approach as compared to using truncated SVD.

QIAN [4] has conducted the analysis of two information retrieval techniques; one is a random mapping (RM) and the other is latent semantic Indexing (LSI). His study is based on theoretical analysis of both techniques to address the issues of self-organizing feature map neural network within ahigh dimens ional environment of text analysis and problem associated with input vector spaces. Their research contributed towards a solution for such issues with text processing which is RM based fast latent semantic indexing method. Fast LSI is cost effective and can be front-end for Kohonen SOFM neural network, which results in high feasibility real-time, accurate classification in a higher dimension text environment. This method of text processing is on its initial stage so can be further improved to solve some other related problem too.

Due to the large size of data which has some sort of exponential property, it is necessary to provide a scalable framework which further capable of indexing and searching web contents i.e. Text, Music, and Images. Only scalability is not the issue associated with information retrieval,l but it should cost effective too. The authors have combined information retrieval and the peer to peer system which gives a more practical look. They also used the search algorithm which is used in peer to peer networks to place document using the LSI method. The authors also have discussed the limitations of LSI. According to authors' contribution, LSI is not good for heterogeneous data and also utilizes more memory and computation time, so produces inferior results when comparing to the Okapi BM25 algorithm. This study is mainly based o0n modification of Latent semantic indexing method by reducing cost of SVD by reducing the input matrix size through clustering and term selection. Their efforts results in a 76% improvement in recall and to improve quality of retrieval used low dimensional sub vectors of semantic vectors for clustering documents which further uses Okapi for search guide and document selection. [5]

Atreya *et al*. [6] compared the retrieval performance of LSI with BM25 benchmark. They have used the TREC 2, 7, 8 and 2004 collections as a data set to perform the evaluation. LSI, which explores the hidden relationship between terms, concepts and documents, is modified to achieve better retrieval accuracy. The performance of LSI was first improved by incorporating Okapi BM25 weights for terms in documents. The second attempt was made by incorporating novel scoring methods, which used score regularization and query expansion in the LSI framework. In the third attempt, LSI framework was combined with BM25 for improved performance. The performance of these modifications of LSI was better than the previous LSI implementation, but they were consistently less accurate then BM25.

The authors claim that the negative results are consistent as they have tried more variants of LSI then the existing work. An understandable reason of low performance by LSI and its variants may be due to a large number of dimensions of data set, but the authors hasve verified that degraded performance is inherent to LSI and large dimension of dataset has no effect on it. Comparing with existing work and moderate dimensions of all four TREC collections it can be concluded that LSI has failed to perform better than the existing benchmark and it is not due to just dimensionality.

Clarizia *et al*. [7] explains the need of textual analysis as it use the statistical techniques to systematically explore the structure of text documents and associate to the original text for suitable interpretations. The authors applied topic modeling, which is a probabilistic model of text analysis, for analysis of corpus of text documents about electromagnetic pollution. The proposed methodology reveals the relationship of documents with its word-frequency. The meanings of a document are distributed to its continuum, which indicates that the inference of the meanings can be made by performing multilevel analysis of the document. They constructed a comprehensive and synthetic representation of a corpus of documents taken from WHO (World Health Organization).They have constructed ontology to represent documents and to extract the required terms of the spectrum of word-frequencies. The authors proposed an application of this technique by representing a document with an ontology, which will enable to store fewer meaningful words instead of the whole document with an advantage of storage space and retrieval time.

Lin *et al*. [8] has designed a text-mining application for the Parliamentary library of Taiwan. The public intended to keep track of each legislator's performance by tracking their daily activities and inquiries at the Legislative Yuan. The designed tool will automatically classify the documents denoting to each legislator and then depicted their performance with corresponding proportions on certain categories.

The authors first created a basic categorical structure of legislative Yuan. A two state clustering was performed for feature selection from these documents. The classification model for these documents was based on SVM, which categorize them according to designed categorical structure. The results of the proposed models were demonstrated experimentally and it provided the public with a facility to monitor and track the legislator's activities and performance using the information obtained from Legislative Yuan. The system demonstrated real world application by providing matching results for experts and the general public.

## METHODOLOGY
### A. LINGO
The label Induction grouping algorithm is based on two phases as given below. The pseudo code of LINGO is taken from Osiński's work. [1]

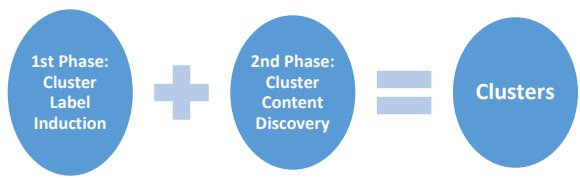

**Figure 2: Phases of LINGO**

Contents assignment is based on a Vector space model which is a well-known and successful IR technique. VSM is based on literal matching not uses the idea of synonymy and it uses on word input. It assigns contents based on word matching which means if a cluster label has that term, then it will assign else ignores the document. We have used Latent Semantic Analysis technique in replacement of VSM to assign content to the Label which is proved through experimental results. Secondly, to address the limitation of the new modified algorithm by analyzing feasibility of using LSI+Okapi BM25 which provide more cost effective results and also benefits of using LSI by retaining concept of search. [9]

### B. Latent Semantic Indexing Method
There are a lot of limitations if we use matching techniques based on simple terms. The main function of latent semantic indexing is to provide the relationships and associations between documents, terms and concept. LSI has statistical background and so uses statistically derived concepts in contrary to the vector space model which only relies on term matching. LSI is mainly associated with term of synonymy which means it accounts for concepts too not only term matching. The core of LSI is Singular value decomposition, it uses truncated SVD for the transformation of the high dimensional document into a lower dimensional semantic vector. Basically LSI is mathematical or statistical technique used for extraction and interpretation of contextual usage of text in a paragraph\passage. To apply LSI, we need to represent text as a matrix having rows as unique words and column would be the text. Now the every cell of the matrix contains the frequency associated with a given word in the text paragraph. Frequency is further utilized to apply a weighted function and will result in the importance of the word in the text. After applying weighted function Singular value decomposition is applied on matrix. Now coming up to working of SVD, this is initiated through building a rectangular matrix into a product of three different matrices. All three matrices provide different values such as the first one is the orthogonal factor values vector of elements, the other is of same original entities and the last one is a diagonal matrix. LSI works on input as a text with inconsistent word choice and word usage is estimated by truncated SVD.

Assuming that rank of term document matrix A is r then by applying SVD, the product of three matrices is produced .
$$A = U\Sigma V^T$$
In this formula, U = (u1, . . . , ur) ∈ Rt×r, Σ = diag(σ_1, . . . , σ_r) ∈ Rr×r, and
V = (v1, . . . , vr) ∈ Rd×r. V T
Here U and V both are column-orthonormal.
$$A_K = U_K \Sigma_K V_K^T$$
Now LSI results in approximation of A having a rank-K matrix for K largest singular values,
Uk = (u1, . . . , uk), Σk = diag(σ_1, . . . , σ_k), Vk = (v1, . . . , vk).
Row i of Uk ∈ Rt×k is the representation of term i in the k-dimensional semantic space.
$$q = U_K^T q$$
$$q = \sum_K^{-1} U_K^T q$$
By using above two equations, document d vector q ∈ Rt×1 can be folded into the k-dimensional semantic space.

**Figure 3: Pseudo code for modified LINGO algorithm**

```
1: D ← input documents (or snippets)
      {STEP 1: Preprocessing}
2: for all d ∈ D do
3:    perform test segmentation of d;
4:    if language of d recognized then
5:        apply stemming and mark stop-words in d;
6:    end if
7: end for
      {STEP 2: Frequent phrase Extraction}
8: concatenate all documents;
9: P_c ← discover complete phrases;
10: P_f ← p: {p belongs P_c ^ frequency (p) > Term Frequency Threshold};
      {STEP 3: Cluster label Induction}
11: A ← term-document matrix of terms marked as stop-words and with frequency
        higher than the Term Frequency Threshold;
12: Σ, U, V ← SVD (A); {Product of SVD decomposition of A}
13: k ← 0; {start with zero clusters}
14: n ← rank (A);
15: repeat
16:    k ← k+1;
17:    q ← ( Σ_{i=1}^{k} Σii) / (Σ_{i=1}^{n} Σii );
18: until q < Candidate Label Threshold;
19: P ← phrase matrix for P_f;
20: for all columns of U_k^T P do
21:    find the largest component m_i in the column;
22:    add the corresponding phrase to the Cluster Label Candidates set;
23:    Label Score ← m_i;
24: end for
25: calculate cosine similarities between all pairs of candidate labels;
26: identify groups of labels that exceed the Label Similarity Threshold;
27: for all groups of similar labels do
28:    select one label with the highest score;
29: end for
      {STEP 4: Cluster Content Discovery}
30: A ← term-document matrix of terms marked as stop-words and with frequency
        higher than the Term Frequency Threshold;
31: Σ, U, V ← SVD (A); {Product of SVD decomposition of A}
32: U_k ← reduced U matrix obtained from SVD (A)
33: Σ_k ← reduced sigma matrix of singular values
34: V_k^T ← reduced matrix of left singular vectors
35: A_k = U_k Σ_k V_k^T ← reduced t x d matrix upto k terms to reduce noise.
32: for all L ∈ Cluster Label Candidates do
31:    create cluster C = Q^T * A_k described with L;
32:    add to C all documents whose similarity
        to C exceeds the Snippet Assignment Threshold;
33: end for
34: put all unassigned documents in the "Others" group;
      {STEP 5: Final Cluster Formation}
35: for all clusters do
36:    cluster Score ← labelScores * ||C||;
37: end for
```

### C. Okapi BM25

Okapi BM25 [9], is modern probabilistic algorithm and also best vector based scoring algorithm too. SO in LSI, Scoring is also part of the technique, but our aim is to get more accurate results even if the dataset is very large which the limitation of LSI is. So by using BM25 weighting method as an alternative to log-entropy with LSI provides better results as IDF factor of BM25 has both desirable properties. When we use BM25 with LSI, it gives negative impact to the documents for the term appearing more than half of the documents. Here is the formula for scoring document d. [6] [9]

$$\sum_i IDF(q_i) \frac{count(q_i, d)(k_1 + 1)}{count(q_i, d) + k_1(1 - b + b \cdot l(d)/L)}$$

The factor IDF (t) for term t is

$$IDF(t) = \frac{\log(n - n(t) + 0.5)}{n(t) + 0.5}$$

> $q_i$: the *ith* query term,
> $k_1$: is a parameter for scaling of term frequency.
> *b:* is a parameter for scaling of document length.
> *Count* ($q_i$, d) is the frequency of term $q_i$ in d.
> l(d) : length of document d.
> L: average document length.

When using BM25 for scoring, the main difference in score is due to non- normalized behavior of column of BM25 matrix which is normalized in case of cosine similarity. Conceptual reason of this is due to BM25 is not geometric and the practical perspective is that b already results in approximate length normalization.

## RESULTS and DISCUSSION

Research includes application of algorithm on five different datasets which have approximately 115 documents and on average twenty labels. Results are documented as below for LINGO algorithm and when using latent semantic indexing for content discovery in phase two of LINGO. Table 1 to table 10 contains the results when both algorithms are applied.

Table 1: Results of Dataset 1 using LINGO

| Serial # | No of Doc's in the cluster | Score of the cluster | Doc's of the cluster |
|---|---|---|---|
| Cluster 1 | 13 | 5.34E+00 | 14,24,26,29,35,37,47,55,67,80,87,93,114, |
| Cluster 2 | 7 | 3.97E+00 | 1,22,35,37,80,90,98 |
| Cluster 3 | 12 | 4.25E+00 | 1,8,15,33,51,52,68,80,85,94,108,109, |
| Cluster 4 | 12 | 6.87E+00 | 14,16,17,18,43,51,55,76,92,108,110, |
| Cluster 5 | 4 | 1.24E+00 | 3,61,90,93, |
| Cluster 6 | 14 | 4.40E+00 | 4,9,21,24,25,27,30,32,65,79,85,100,102,103, |
| Cluster 7 | 4 | 1.48E+00 | 44,68,84,107 |
| Cluster 8 | 8 | 2.55E+00 | 17,18,51,76,92,96,108,110, |
| Cluster 9 | 10 | 5.24E+00 | 1,8,16,41,51,52,68,80,85,108, |
| Cluster 10 | 4 | 1.64E+00 | 3,61,90,93 |
| Cluster 11 | 5 | 2.22E+00 | 46,61,73,84,93 |
| Cluster 12 | 5 | 1.77E+00 | 2,6,11,38,64, |
| Cluster 13 | 8 | 2.11E+00 | 20,40,62,77,79,84,106,111 |
| Cluster 14 | 6 | 2.41E+00 | 1,14,42,45,55,111, |
| Cluster 15 | 4 | 1.25E+00 | 6,64,94,111 |
| Cluster 16 | 7 | 2.18E+00 | 4,28,41,59,77,80,106 |
| Cluster 17 | 4 | 1.17E+00 | 22,35,37,98, |
| Cluster 18 | 2 | 4.90E-01 | 1,37, |
| Cluster 19 | 9 | 3.43E+00 | 1,8,33,51,52,68,80,85,108, |
| Cluster 20 | 11 | 4.31E+00 | 1,8,15,33,51,52,68,85,94,108,109, |
| Cluster 21 | 3 | 7.11E-01 | 19,49,84, |
| Other topics = | 42 | | 5,7,10,12,13,23,31,34,36,48,50,53,54,56,57,58,60,63,66,69,70,71,72,74,75,78,81,82,83,86,88,89,91,95,97,99,101,104,105,112,113,115 |

Table 2: Results of Dataset 1 using LSI

| Serial # | No of Doc's in the cluster | Score of the cluster | Doc's of the cluster |
|---|---|---|---|
| Cluster 1 | 17 | 6.98E+00 | 3,14,16,24,26,29,43,45,47,55,67,77,80,82,87,93,114, |
| Cluster 2 | 23 | 1.30E+01 | 1,2,3,10,13,22,25,35,37,38,53,60,67,76,79,83,90,91,98,102,106,110,115, |
| Cluster 3 | 25 | 8.86E+00 | 1,8,15,18,33,41,44,47,48,51,52,58,68,76,80,81,85,86,94,96,100,107,108,109,112, |
| Cluster 4 | 15 | 8.59E+00 | 14,16,17,18,33,43,49,51,55,76,92,96,98,108,110, |
| Cluster 5 | 8 | 2.49E+00 | 3,46,61,67,73,90,91,93, |
| Cluster 6 | 17 | 5.34E+00 | 1,4,9,21,24,25,26,27,30,32,58,65,79,85,100,102,103, |
| Cluster 7 | 7 | 2.59E+00 | 1,44,52,68,84,101,107, |
| Cluster 8 | 11 | 3.51E+00 | 17,18,33,49,51,52,76,92,96,108,110, |
| Cluster 9 | 10 | 5.24E+00 | 8,33,47,51,52,80,85,87,108,109, |
| Cluster 10 | 8 | 3.27E+00 | 3,46,61,67,73,90,91,93, |
| Cluster 11 | 9 | 3.99E+00 | 46,60,61,72,73,84,90,93,97, |
| Cluster 12 | 9 | 3.19E+00 | 2,6,11,38,64,91,94,95,104, |
| Cluster 13 | 10 | 2.64E+00 | 19,20,40,62,77,79,82,88,103,106, |
| Cluster 14 | 7 | 2.81E+00 | 1,14,42,43,55,69,111, |
| Cluster 15 | 5 | 1.56E+00 | 6,22,53,64,94, |
| Cluster 16 | 10 | 3.12E+00 | 28,41,56,59,77,80,82,87,106,113, |
| Cluster 17 | 19 | 5.58E+00 | 2,3,10,13,22,25,35,37,38,60,67,76,79,83,90,91,98,110,115, |
| Cluster 18 | 5 | 1.23E+00 | 3,35,37,90,98, |
| Cluster 19 | 12 | 4.57E+00 | 1,8,15,33,47,51,52,68,80,85,108,109, |
| Cluster 20 | 12 | 4.71E+00 | 1,8,15,33,51,52,68,85,94,96,108,109, |
| Cluster 21 | 3 | 7.11E-01 | 17,19,84, |
| Other topics = | 21 | | 5,7,12,23,31,34,36,39,50,54,57,63,66,70,71,74,75,78,89,99,105 |

Table 3: Results of Dataset 2 using LINGO

| Serial # | No of Doc's in the cluster | Score of the cluster | Doc's of the cluster |
|---|---|---|---|
| Cluster 1 | 17 | 7.01E+00 | 16,27,38,40,42,47,50,64,65,66,71,77,79,80,91,93,94, |
| Cluster 2 | 2 | 7.75E-01 | 56,85, |
| Cluster 3 | 18 | 7.53E+00 | 7,8,15,17,18,20,25,26,29,33,40,46,54,55,64,65,76,82, |
| Cluster 4 | 6 | 2.61E+00 | 32,33,42,65,73,77, |
| Cluster 5 | 5 | 1.87E+00 | 1,24,34,35,45, |
| Cluster 6 | 3 | 9.24E-01 | 66,70,85, |
| Cluster 7 | 3 | 8.82E-01 | 66,70,85, |
| Cluster 8 | 4 | 1.38E+00 | 7,12,18,36, |
| Cluster 9 | 3 | 1.14E+00 | 7,12,18,36, |
| Cluster 10 | 3 | 1.57E+00 | 66,73,94, |
| Cluster 11 | 2 | 6.18E-01 | 16,88, |
| Cluster 12 | 2 | 7.95E-01 | 60,90, |
| Cluster 13 | 8 | 3.02E+00 | 15,26,28,54,59,72,89,96, |
| Cluster 14 | 2 | 6.45E-01 | 33,63, |
| Cluster 15 | 4 | 1.37E+00 | 17,22,23,47, |
| Cluster 16 | 2 | 1.01E+00 | 32,36, |
| Cluster 17 | 3 | 2.18E+00 | 28,44,59, |
| Cluster 18 | 2 | 5.80E-01 | 38,47, |
| Cluster 19 | 5 | 1.50E+00 | 12,64,79,80,92, |
| Cluster 20 | 5 | 1.58E+00 | 23,54,59,60,72, |
| Other topics = | 55 | | 2,3,4,5,6,9,10,11,13,14,19,30,31,37,39,41,43,48,49,51,52,53,57,58,61,62,68,69,74,75,78,81,83,84,86,87,97,98,99,100,101,102,103,104,105,106,107,108,109,110,111,112,113,114,115 |

Table 4: Results of Dataset 2 using LSI

| Serial # | No of Doc's in the cluster | Score of the cluster | Doc's of the cluster |
|---|---|---|---|
| Cluster 1 | 19 | 7.83E+00 | 16,17,22,38,40,42,47,64,65,66,71,73,77,79,80,91,92,93,94, |
| Cluster 2 | 4 | 1.55E+00 | 4,51,66,85, |
| Cluster 3 | 23 | 9.63E+00 | 7,8,15,17,18,20,25,26,29,33,37,40,41,46,54,55,64,65,76,77,82,87,95, |
| Cluster 4 | 11 | 4.78E+00 | 32,33,36,42,64,65,71,73,77,92,94, |
| Cluster 5 | 9 | 3.37E+00 | 1,24,30,34,35,40,42,45,68, |
| Cluster 6 | 2 | 6.16E-01 | 66,85, |
| Cluster 7 | 9 | 2.65E+00 | 26,27,50,54,61,67,70,71,95, |
| Cluster 8 | 9 | 3.12E+00 | 7,12,18,19,36,60,68,80,92, |
| Cluster 9 | 13 | 4.94E+00 | 21,27,41,45,48,50,53,67,69,74,84,93,95, |
| Cluster 10 | 6 | 3.15E+00 | 16,65,66,73,77,94, |
| Cluster 11 | 5 | 1.55E+00 | 10,16,30,79,88, |
| Cluster 12 | 7 | 2.78E+00 | 7,20,60,71,87,90,91, |
| Cluster 13 | 8 | 3.02E+00 | 15,26,28,54,59,66,72,89, |
| Cluster 14 | 2 | 6.45E-01 | 32,63, |
| Cluster 15 | 8 | 2.73E+00 | 17,22,23,38,47,48,57,80, |
| Cluster 16 | 2 | 1.01E+00 | 32,36, |
| Cluster 17 | 5 | 3.63E+00 | 28,59,63,72,89, |
| Cluster 18 | 3 | 8.70E-01 | 17,22,47, |
| Cluster 19 | 12 | 3.60E+00 | 7,12,16,18,36,41,64,68,69,79,80,92, |
| Cluster 20 | 9 | 2.84E+00 | 7,23,28,54,59,60,68,72,89, |
| Other topics = | 42 | | 2,3,5,6,9,11,13,14,31,39,43,44,49,52,56,58,62,75,78,81,83,86,96,97,98,99,100,101,102,103,104,105,106,107,108,109,110,111,112,113,114,115 |

Table 5: Results of Dataset 3 using LINGO

| Serial # | No of Doc's in the cluster | Score of the cluster | Doc's of the cluster |
|---|---|---|---|
| Cluster 1 | 9 | 5.19E+00 | 1,3,17,20,26,63,88,89,92, |
| Cluster 2 | 8 | 3.44E+00 | 9,23,28,41,48,51,71,93, |
| Cluster 3 | 10 | 4.16E+00 | 1,3,17,20,21,26,63,88,89,92, |
| Cluster 4 | 7 | 2.46E+00 | 5,6,9,28,56,69,91, |
| Cluster 5 | 5 | 2.06E+00 | 3,11,56,86,89, |
| Cluster 6 | 9 | 3.10E+00 | 1,9,11,14,31,33,59,67,73, |
| Cluster 7 | 3 | 1.38E+00 | 70,78,84, |
| Cluster 8 | 2 | 8.81E-01 | 69,88, |
| Cluster 9 | 5 | 1.70E+00 | 32,61,80,83,93, |
| Cluster 10 | 5 | 1.95E+00 | 17,61,83,87,93, |
| Cluster 11 | 2 | 8.13E-01 | 57,72, |
| Cluster 12 | 5 | 1.88E+00 | 61,82,83,93,94, |
| Cluster 13 | 9 | 4.72E+00 | 22,26,27,36,38,39,43,61,85, |
| Cluster 14 | 3 | 2.17E+00 | 46,50,63, |
| Cluster 15 | 9 | 5.37E+00 | 1,9,11,14,31,33,59,67,73, |
| Cluster 16 | 2 | 9.84E-01 | 1,52, |
| Cluster 17 | 4 | 2.17E+00 | 39,58,76,84, |
| Cluster 18 | 4 | 1.23E+00 | 45,51,68,73, |
| Cluster 19 | 2 | 8.37E-01 | 77,79, |
| Cluster 20 | 2 | 7.65E-01 | 74,79, |
| Other topics = | 55 | | 2,4,7,8,10,12,13,15,16,18,19,24,25,29,30,34,35,37,40,42,44,47,49,53,54,55,60,62,64,65,66,75,81,90,95,96,97,98,99,100,101,102,103,104,105,106,107,108,10 |

**Table 6: Results of Dataset 3 using LSI**

| Serial # | No of Doc's in the cluster | Score of the cluster | Doc's of the cluster |
|---|---|---|---|
| Cluster 1 | 15 | 8.64E+00 | 1,3,11,17,20,21,26,49,56,63,66,86,88,89,92, |
| Cluster 2 | 8 | 3.44E+00 | 9,14,28,41,48,69,71,93, |
| Cluster 3 | 17 | 7.08E+00 | 1,3,11,17,20,21,26,49,51,56,63,66,69,86,88,89,92, |
| Cluster 4 | 8 | 2.82E+00 | 5,6,9,28,41,69,75,91, |
| Cluster 5 | 13 | 5.37E+00 | 3,11,13,17,26,30,53,56,63,65,86,89, |
| Cluster 6 | 14 | 4.82E+00 | 1,9,11,14,16,18,28,31,33,40,56,59,67,73, |
| Cluster 7 | 7 | 3.23E+00 | 42,58,70,76,78,84,87, |
| Cluster 8 | 9 | 3.97E+00 | 20,25,28,41,69,75,88,89,92, |
| Cluster 9 | 17 | 5.79E+00 | 22,30,32,38,43,49,61,69,75,80,81,82,83,85,92,93,94, |
| Cluster 10 | 7 | 2.73E+00 | 17,32,42,70,87,92,93, |
| Cluster 11 | 3 | 1.22E+00 | 23,57,72, |
| Cluster 12 | 26 | 9.78E+00 | 15,20,21,28,30,32,34,36,37,38,39,44,53,60,61,62,65,73,79,80,82,83,86,90,93,94 |
| Cluster 13 | 9 | 4.72E+00 | 22,26,27,36,38,39,43,61,85, |
| Cluster 14 | 7 | 5.06E+00 | 29,33,46,50,63,77,90, |
| Cluster 15 | 8 | 4.78E+00 | 9,11,14,16,31,33,59,67, |
| Cluster 16 | 2 | 9.84E-01 | 35,52, |
| Cluster 17 | 5 | 2.72E+00 | 1,39,58,76,84, |
| Cluster 18 | 9 | 2.78E+00 | 2,30,35,45,51,52,66,68,73, |
| Cluster 19 | 4 | 1.67E+00 | 60,77,79,83, |
| Cluster 20 | 1 | 3.83E-01 | 79, |
| Other topics = | 33 | | 4,7,8,10,12,19,24,47,54,55,64,74,95,96,97,98,99,100,101,102,103,104,105,106,107,108,109,110,111,112,113,114,115 |

**Table 7: Results of Dataset 4 using LINGO**

| Serial # | No of Doc's in the cluster | Score of the cluster | Doc's of the cluster |
|---|---|---|---|
| Cluster 1 | 4 | 1.34E+00 | 28,39,67,104, |
| Cluster 2 | 7 | 5.03E+00 | 6,12,14,20,34,44,76, |
| Cluster 3 | 7 | 1.97E+00 | 13,32,39,61,64,83,98, |
| Cluster 4 | 3 | 1.42E+00 | 36,40,59, |
| Cluster 5 | 2 | 5.93E-01 | 6,93, |
| Cluster 6 | 5 | 1.44E+00 | 18,24,43,66,81, |
| Cluster 7 | 3 | 1.11E+00 | 15,46,62, |
| Cluster 8 | 3 | 9.52E-01 | 47,57,75, |
| Cluster 9 | 2 | 5.61E-01 | 65,95, |
| Cluster 10 | 7 | 2.48E+00 | 25,30,59,76,90,91,105, |
| Cluster 11 | 3 | 1.24E+00 | 34,69,96, |
| Cluster 12 | 3 | 8.69E-01 | 85,91,97, |
| Cluster 13 | 2 | 6.03E-01 | 77,90, |
| Cluster 14 | 4 | 1.08E+00 | 3,10,21,45, |
| Cluster 15 | 2 | 7.41E-01 | 71,104, |
| Cluster 16 | 6 | 1.36E+00 | 47,52,57,60,75,81, |
| Cluster 17 | 7 | 2.62E+00 | 50,58,70,77,80,93,103, |
| Cluster 18 | 3 | 8.84E-01 | 42,74,99, |
| Cluster 19 | 2 | 5.84E-01 | 30,101, |
| Cluster 20 | 2 | 5.05E-01 | 53,89, |
| Cluster 21 | 7 | 1.82E+00 | 1,4,38,41,49,56,92, |
| Other topics = | 46 | | 2,5,7,8,9,11,16,17,19,22,23,26,27,29,31,33,35,37,48,51,54,55,63,68,72,73,78,79,82,84,86,87,88,94,100,102,106,107,108,109,110,111,112,113,114,115 |

**Table 8: Results of Dataset 4 using LSI**

| Serial # | No of Doc's in the cluster | Score of the cluster | Doc's of the cluster |
|---|---|---|---|
| Cluster 1 | 11 | 3.69E+00 | 4,16,18,23,25,28,33,39,67,71,104, |
| Cluster 2 | 10 | 7.19E+00 | 6,12,14,19,20,34,40,44,69,93, |
| Cluster 3 | 9 | 2.53E+00 | 13,32,39,61,64,65,83,97,98, |
| Cluster 4 | 7 | 3.32E+00 | 25,36,40,44,59,76,105, |
| Cluster 5 | 4 | 1.19E+00 | 3,6,21,93, |
| Cluster 6 | 11 | 3.16E+00 | 11,17,18,24,33,43,63,66,68,73,81, |
| Cluster 7 | 6 | 2.21E+00 | 15,17,46,62,72,102, |
| Cluster 8 | 7 | 2.22E+00 | 2,47,52,57,60,75,81, |
| Cluster 9 | 4 | 1.12E+00 | 13,65,95,97, |
| Cluster 10 | 9 | 3.19E+00 | 25,30,36,59,76,90,91,101,105, |
| Cluster 11 | 6 | 2.48E+00 | 12,30,34,69,71,87, |
| Cluster 12 | 6 | 1.74E+00 | 13,36,65,85,91,97, |
| Cluster 13 | 5 | 1.51E+00 | 50,73,77,80,90, |
| Cluster 14 | 6 | 1.62E+00 | 3,5,6,10,21,45, |
| Cluster 15 | 6 | 2.22E+00 | 28,39,67,69,71,104, |
| Cluster 16 | 7 | 1.59E+00 | 2,47,52,57,60,75,81, |
| Cluster 17 | 13 | 4.87E+00 | 7,50,58,70,73,77,78,80,90,92,93,101,103, |
| Cluster 18 | 5 | 1.47E+00 | 42,61,74,82,99, |
| Cluster 19 | 8 | 2.34E+00 | 22,30,70,92,100,101,103,105, |
| Cluster 20 | 6 | 1.51E+00 | 8,36,53,85,89,100, |
| Cluster 21 | 11 | 2.86E+00 | 1,4,38,41,49,51,56,84,86,87,92, |
| Other topics = | 24 | | 9,26,27,29,31,35,37,48,54,55,79,88,94,96,106,107,108,109,110,111,112,113,114,115 |

**Table 9: Results of Dataset 5 using LINGO**

| Serial # | No of Doc's In the cluster | Score of the cluster | Doc's of the cluster |
|---|---|---|---|
| Cluster 1 | 4 | 1.40E+00 | 64,75,105,115, |
| Cluster 2 | 5 | 2.04E+00 | 32,51,84,102,108, |
| Cluster 3 | 16 | 1.27E+01 | 12,19,22,24,47,55,60,70,77,78,97,98,103,106,112,116, |
| Cluster 4 | 12 | 3.54E+00 | 4,9,12,22,27,34,37,40,41,67,73,107, |
| Cluster 5 | 5 | 4.30E+00 | 22,24,37,48,59, |
| Cluster 6 | 9 | 2.80E+00 | 16,21,33,34,39,41,92,96,114, |
| Cluster 7 | 5 | 2.21E+00 | 36,51,82,98,106, |
| Cluster 8 | 7 | 2.87E+00 | 28,62,66,68,75,90,110, |
| Cluster 9 | 6 | 1.91E+00 | 1,3,8,52,83,115, |
| Cluster 10 | 3 | 8.16E-01 | 4,34,53, |
| Cluster 11 | 5 | 1.29E+00 | 18,29,65,88,113, |
| Cluster 12 | 3 | 8.79E-01 | 10,39,51, |
| Cluster 13 | 4 | 1.04E+00 | 24,28,40,110, |
| Cluster 14 | 3 | 7.83E-01 | 20,65,71, |
| Cluster 15 | 2 | 4.38E-01 | 39,73, |
| Cluster 16 | 3 | 7.79E-01 | 21,28,111, |
| Cluster 17 | 2 | 4.20E-01 | 25,86, |
| Cluster 18 | 2 | 5.41E-01 | 73,81, |
| Cluster 19 | 2 | 4.15E-01 | 14,87, |
| Cluster 20 | 2 | 9.23E-01 | 2,7, |
| Cluster 21 | 3 | 1.00E+00 | 98,104,110, |
| Other topics = | 41 | | 5,6,11,13,15,17,23,26,30,31,35,38,42,43,44,45,46,49,50,54,56,57,58,61,63,69,72,74,76,79,80,85,89,91,93,94,95,99,100,101,109 |

**Table 10: Results of Dataset 5 using LSI**

| Serial # | No of Doc's in the cluster | Score of the cluster | Doc's of the cluster |
|---|---|---|---|
| Cluster 1 | 10 | 3.49E+00 | 23,28,55,64,66,68,75,89,105,115, |
| Cluster 2 | 7 | 2.85E+00 | 32,51,57,84,102,108,120, |
| Cluster 3 | 22 | 1.75E+01 | 2,12,13,19,24,28,30,43,47,50,55,60,70,77,78,97,98,103,106,108,112,116, |
| Cluster 4 | 17 | 5.01E+00 | 4,9,22,24,27,34,40,41,48,53,59,67,73,74,80,96,107, |
| Cluster 5 | 8 | 6.87E+00 | 9,22,24,27,40,48,59,107, |
| Cluster 6 | 11 | 3.42E+00 | 4,16,21,33,34,39,41,53,92,96,114, |
| Cluster 7 | 9 | 3.97E+00 | 5,26,36,51,82,97,98,103,106, |
| Cluster 8 | 10 | 4.10E+00 | 17,28,48,62,66,68,75,90,105,110, |
| Cluster 9 | 8 | 2.55E+00 | 1,3,8,52,55,83,89,115, |
| Cluster 10 | 6 | 1.63E+00 | 4,16,34,41,53,96, |
| Cluster 11 | 12 | 3.09E+00 | 5,18,20,28,29,63,65,71,85,88,93,113, |
| Cluster 12 | 4 | 8.89E+00 | 10,39,51,52 |
| Cluster 13 | 6 | 1.56E+00 | 9,24,27,28,40,59, |
| Cluster 14 | 8 | 2.09E+00 | 12,18,20,47,52,65,71,84, |
| Cluster 15 | 5 | 1.09E+00 | 33,39,73,81,85, |
| Cluster 16 | 7 | 1.82E+00 | 21,28,37,63,93,111,113, |
| Cluster 17 | 3 | 6.29E-01 | 54,86,117, |
| Cluster 18 | 1 | 2.70E-01 | 73, |
| Cluster 19 | 6 | 1.25E+00 | 14,15,42,47,87,91, |
| Cluster 20 | 4 | 1.85E+00 | 2,7,77,108, |
| Cluster 21 | 3 | 1.70E+00 | 98,104,110, |
| Other topics = | 25 | | 6,10,11,25,31,35,38,44,45,46,49,56,58,61,69,72,76,79,94,95,99,100,101,104,109 |

**Graphical view of results**

The results we obtained from both LINGO and LSI are graphically represented in Graph 1. This gives the complete picture of this comparison between LINGO original method and LINGO when modified through using LSI for cluster content discovery. From graphical view it is clear that Modified LINGO provides better results so using LSI for content discover of cluster is practically better than the existing method.

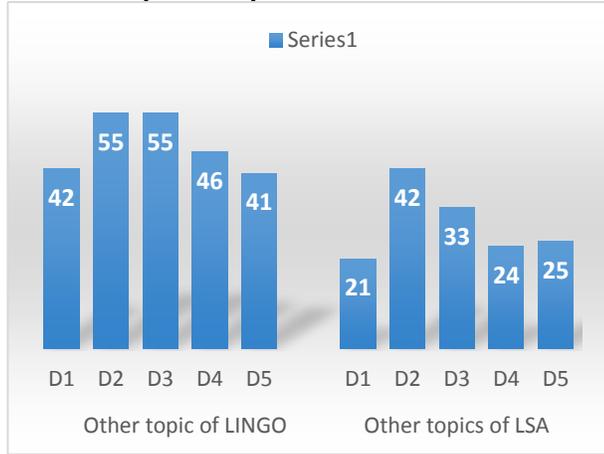

Graph 1: Comparison between LINGO & LSI

## Quality of Clusters Using LINGO and LSI: graphical view

Now for the computation of Cluster Score, we use the formula as below

*Cluster score = score of label x no of documents in the cluster*

By applying this formula on our datasets, we may achieve the quality clusters for search results. Following graph shows the results of score of LING and LINGO with using LSI.

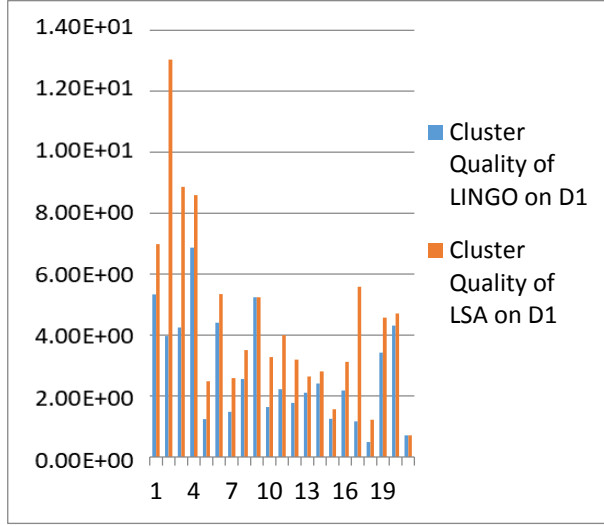

Graph 2: Cluster Quality for Dataset using LINGO & LSI

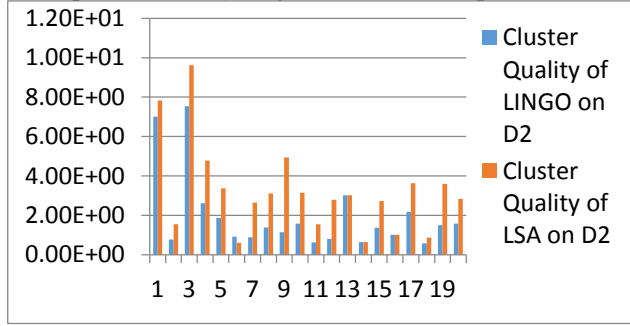

Graph 3: Cluster Quality for Dataset 2 using LINGO & LSI

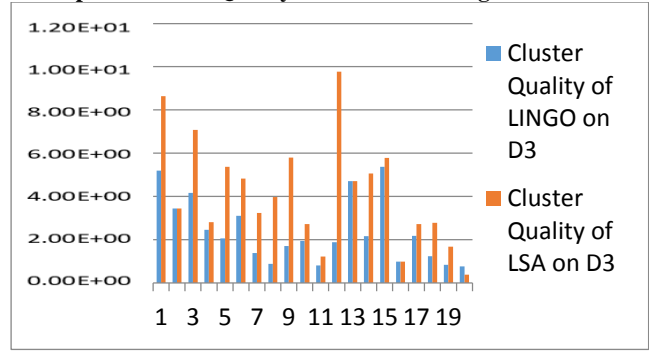

Graph 4: Cluster Quality for Dataset 3 using LINGO & LSI

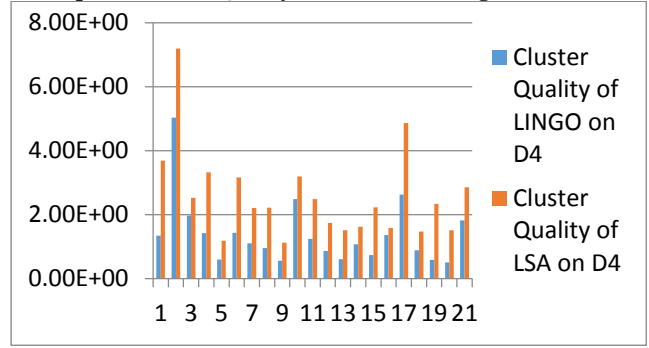

Graph 5: Cluster Quality for Dataset 4 using LINGO & LSI

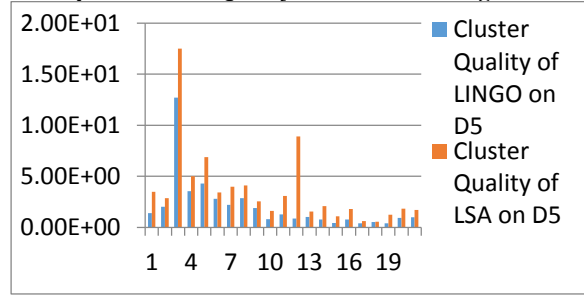

Graph 6: Cluster Quality for Dataset 5 using LINGO & LSI

## CONCLUSION

From results of experiments, it is clear that Latent semantic indexing technique produces better results while comparing with that of vector space model. The fact is that LSI does not limit to small snippets as LINGO does also gives better clusters through assigning better membership of documents. Our research is based on semantic based clustering of web search results and to use the latent semantic indexing for content discovery to add quality of synonymy and it results better when compares to existing LINGO method. New modification results in 40-50% better results than original method. Furthermore, there are few limitation of LSI regarding scalability and cost of implementation. The Studies shows that performance of LSI is significantly improved when LSI uses BM25 as scoring method. So By implementing Okapi BM25 for scoring method in LSI then to apply it on cluster content discovery phase of LINGO results in more improved results. Both modifications in LINGO algorithm are new as far as our knowledge.

**FUTURE WORK**

Future work is based on implementing LINGO for same datasets with using Okapi BM25 in scoring method part of LSI and then to compare results of both. The algorithm can be further improved by using good preprocessing of data as it matters a lot in search performance, quality of labels may also be improved by using pruning technique such as induce hierarchical relationships between the topics.